# Tapping-mode SQUID-on-tip Microscopy with Proximity Josephson Junctions

Matthijs Rog[1], Tycho J. Blom[1], Daan B. Boltje[2], Jimi D. de Haan[2], Remko Fermin[1,3], Jiasen Niu[1,4,6], Yasmin C. Doedes[1], Milan P. Allan[1,2,4,5,6], Kaveh Lahabi[1,2*]

*1 Huygens-Kamerlingh Onnes Laboratory, Leiden University, 2300 RA Leiden, The Netherlands.*
*2 QuantaMap B.V., Robert Boyleweg 4, 2333 CG Leiden, The Netherlands.*
*3 Department of Materials Science & Metallurgy, University of Cambridge, 27 Charles Babbage Road, Cambridge CB3 0FS, United Kingdom*
*4 Faculty of Physics, Ludwig-Maximilians-University Munich, Munich 80799, Germany*
*5 Center for Nano Science (CeNS), Ludwig-Maximilians-University Munich, Munich 80799, Germany*
*6 Munich Center for Quantum Science and Technology (MCQST), Ludwig-Maximilians-University Munich, Munich 80799, Germany*

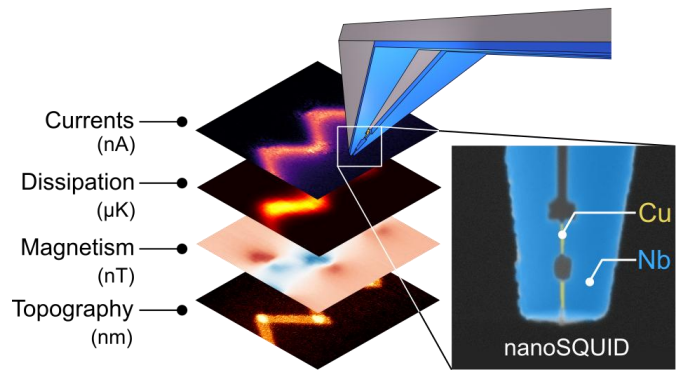

Studying nanoscale dynamics is essential for understanding quantum materials and advancing quantum chip manufacturing. Still, it remains a major challenge to measure non-equilibrium properties such as current and dissipation, and their relation to structure. Scanning nanoprobes utilizing superconducting quantum interference devices (SQUIDs) are uniquely suited here, due to their unparalleled magnetic and thermal sensitivity. Here, we introduce tapping-mode SQUID-on-tip, which combines atomic force microscopy (AFM) with nanoSQUID sensing. Our probes minimize nanoSQUID–sample distance, provide in-plane magnetic sensitivity, and operate without lasers. Frequency multiplexing enables simultaneous imaging of currents, magnetism, dissipation and topography. The large voltage output of our proximity-junction nanoSQUIDs allows us to resolve nanoscale currents as small as 100 nA using a simple four-probe electronic readout without cryogenic amplification. By capturing local magnetic, thermal, and electronic response without external radiation, our technique offers a powerful non-invasive route to study dynamic phenomena in exotic materials and delicate quantum circuits.

Quantum materials stand out by their unusual non-equilibrium properties. From edge currents in topological devices to Planckian dissipation in strongly-correlated matter, understanding the underlying physics requires local insights into the nature of transport and dissipation.[1–4] However, probing these quantities on the nanoscale is notoriously challenging. Moreover, in strongly-interacting systems, it is well understood that transport and dissipation are strongly interconnected with magnetism and system geometry. Due to the lack of a local probe to measure these fundamental quantities simultaneously, their local dynamics and interactions have remained elusive. Notable examples of this are phase transitions in many-body systems, where it is extremely challenging to identify the mechanism that triggers the transition.[5,6]

The need for such experimental probes is mirrored in the efforts towards working quantum computers. All superconducting and spin qubits are sensitive to local dynamics introduced by defects, such as magnetic impurities, vortices and inhomogeneous supercurrents.[7–11] Identifying such sources of decoherence has been a major roadblock, as their small experimental signatures are only detectable at cryogenic temperatures. In contrast, the techniques used for quantum chip diagnostics are either cryogenically incompatible, or suffer from insufficient sensitivity, poor spatial resolution or invasive readout.

This work presents a microscope designed specifically to address these challenges. Our technique is capable of simultaneously recording magnetism, dissipation, charge transport and topography on the nanoscale. Eliminating common sources of invasiveness, this microscope is innately compatible with fragile quantum systems. By recording in-plane magnetic fields with exceptional sensors, this technique can resolve densely packed, complex current flow at the nanoscale with unprecedented detail and sensitivity. These three aspects of our technique open new possibilities in the research on quantum materials and devices inaccessible to established scanning probe techniques.

The integration of quantum sensors in scanning probe microscopy (SPM) has revolutionized our ability to study quantum phenomena on the nanoscale. For instance, scanning nitrogen-vacancy (NV) center microscopy has rapidly established itself as a standard technique, capable of high-resolution imaging of magnetic phenomena and transport. At present, the sensitivity of this technique allows the mapping of small currents approaching 1 µA.[12] Although originally a room-temperature technique, much effort has gone into making scanning NV cryogenically compatible to study superconducting systems.[13–16] What is lacking are complementary cryogenic probes that have similar or better magnetic sensitivity, can measure dissipation, and do not rely on invasive readout schemes which involve lasers and microwave radiation. Superconducting quantum interference devices (SQUIDs) stand out as the ideal sensors to meet these requirements.

SQUIDs are the most sensitive electronic tools for measuring magnetic signals and are innately compatible with cryogenic temperatures. SQUID-based SPM techniques capitalize on this and have led to monumental breakthroughs in our understanding of quantum materials by probing exotic superconducting states,[18–20] phase transitions[21,22] and quantum Hall states.[23] Of particular interest is SQUID-on-tip (SOT) microscopy, which unlocks exceptional spatial resolution by utilizing a nanoSQUID that directly scans the surface. This on-tip approach has enabled SOT to provide novel insights into vortex dynamics, topological matter, twisted bilayer graphene and electron hydrodynamics.[24–29]
In addition to its exceptional magnetic sensitivity, SOT has a unique capacity for high-resolution thermal imaging at cryogenic temperatures. Its sub-µK thermal sensitivity surpasses other thermal microscopes by orders of magnitude, making it the only technique capable of mapping dissipation below the operation limit of a qubit.[30] Furthermore, SQUIDs utilize a fully electronic readout, which does not rely on external radiation, making SOT non-invasive. The ability to probe local magnetism, dissipation and transport, combined with its cryogenic operation, makes SOT ideal for studying fragile quantum systems.

Despite the great accomplishments of this microscope, there is opportunity for major advances that can realize the ultimate resolution and sensitivity of SOT, and pave the way for it to become a mainstream SPM technique. Specifically, SOT lacks the height feedback required for robust topographic imaging. As a result, SOTs typically scan at about 150-250 nm above the surface.[28,31–33] Scanning closer to the surface greatly increases the risk of destroying the SQUID by crashing into the sample. This restriction on the workable scan height not only reduces the measured signal, but also sets the limit for spatial resolution. This issue is rooted in the design of existing SOT probes, where the SQUID is formed at the apex of a quartz tube.[34–36] Although the quartz tip is glued to a tuning fork, the tip-surface coupling is insufficient for stable topographic feedback.

A path to resolving this issue is to combine SOT with the powerful height control of atomic force microscopy (AFM). One implementation of this is SQUID-on-lever (SOL) microscopy, which fabricates the SQUID on a blunted AFM cantilever with a small protruding tip.[37] Laser-based optical readout of the cantilever deflection enables simultaneous SQUID sensing and topographic imaging. SOL has been used to study magnetic ordering in 2D materials and magnetic inhomogeneities in spin qubits.[38,39] However, the sample-sensor distance remains limited to a 200-300 nm gap by the size of the protruding tip and the use of non-contact AFM. More importantly, the laser used for the AFM readout can induce local heating and spurious interference in the sample, analogous to NV microscopy.[17] This undermines the non-invasive electronic readout of the SQUID.

In this work, we present a SQUID-on-tip microscope that closes the SQUID-to-sample gap and provides robust topographic readout without relying on external radiation. Our probes operate fully electronically, without lasers, making them ideal for cryogenic operation. The unique design of our probe enables tapping-mode AFM, allowing us to scan nanofabricated devices with realistic topographic features. Even after weeks of continuous scanning in this mode, the probes show no sign of degradation. Our probes are equipped with high-performance SQUIDs that utilize proximity Josephson junctions that have exceptionally large voltage output. This allows us to simplify the SQUID readout scheme to a four-wire measurement. Using these unique features, we resolve the distribution of nA currents on the nanoscale in a densely packed nanostructure. The exceptional sensitivity of our SQUIDs, combined with the ability to image magnetism, transport, dissipation and topography simultaneously, makes this technique a powerful tool to investigate quantum materials and devices.

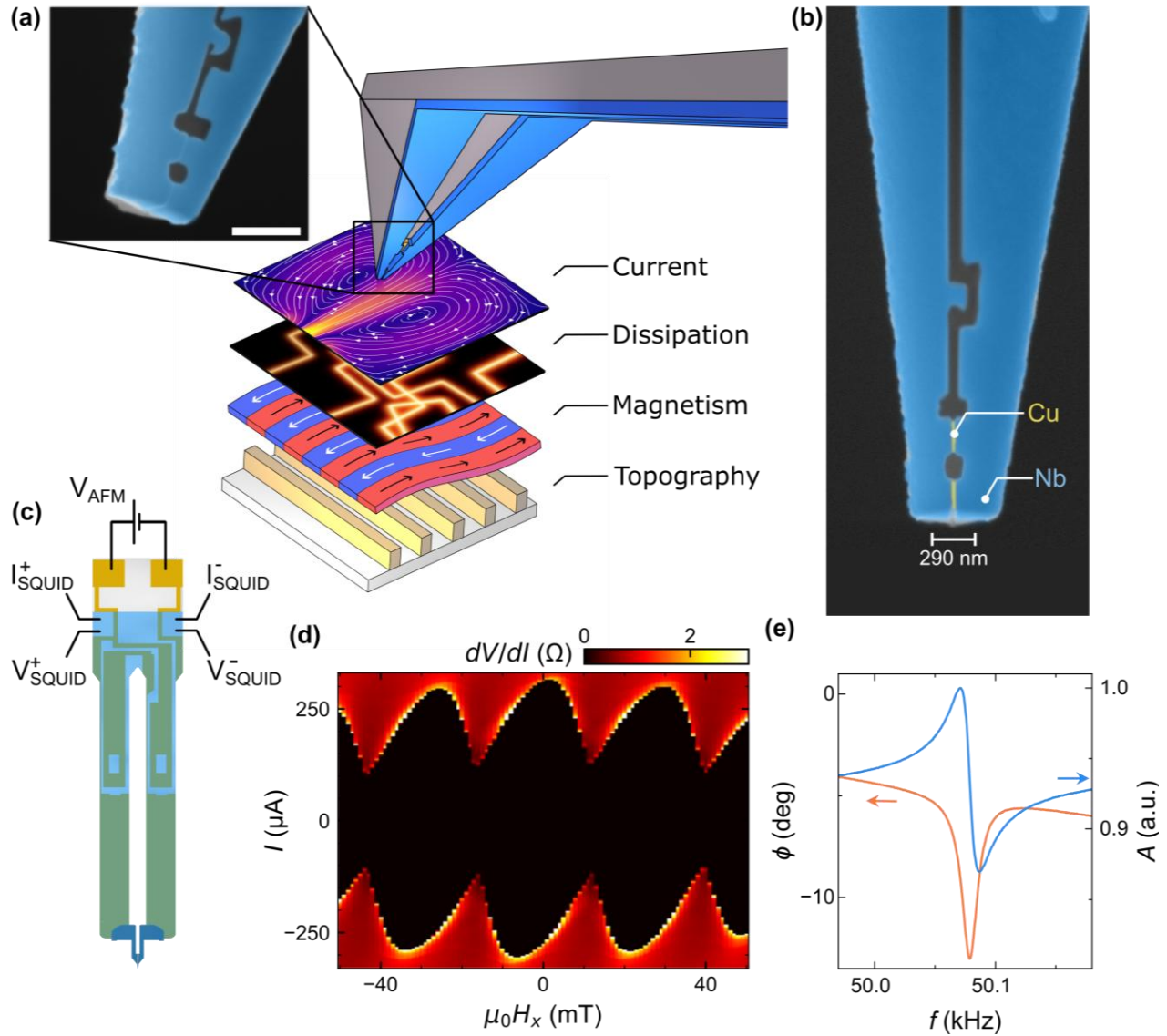

**Figure 1.** Overview of the tapping-mode SQUID-on-tip probe. **(a)** Schematic overview of the microscope. Our probes image current, dissipation, magnetism and topography simultaneously. The inset shows a scanning electron micrograph of a probe with a scalebar of 500 nm. **(b)** A scanning electron micrograph of a SQUID-on-tip probe, taken after seven weeks of continuous usage. The effective diameter of the thin-film SQUID is 290 nm. **(c)** Overview of the quartz tuning fork. The SQUID devices are made on top of a commercial Akiyama tuning fork AFM probe. The tuning fork electrodes are shown in gold. Coverage of the SQUID thin films is shown in blue. **(d)** Supercurrent quantum interference (SQI) pattern of our device recorded at 2.5K. **(e)** Resonance curve of the first cantilever bending mode recorded at 4K. This data was recorded with a SQUID on the probe, illustrating that the presence of the SQUID does not modify the Q-factor of the probe.

We begin by highlighting the key characteristics of our probes. We fabricate our nanoSQUID at the apex of an AFM cantilever strongly coupled to a quartz tuning fork. An overview of the probe is presented in Figures 1a-c. Compared to traditional SOT and SOL probes, ours differ significantly in design and operation. The most apparent distinction is the orientation of the SQUID plane, which has a fixed 62° angle relative to the sample. By choosing this orientation, we allow the SQUID to probe both out-of-plane and in-plane components of the magnetic field. It also provides an additional degree of freedom to field-tune the SQUID, allowing us to maintain an optimal working point during field sweeps. This design is significantly more sensitive to in-plane magnetic signals, which enhances our capacity for spatially resolving small currents (see Supporting Information for a simulation).

We produce our sensors on top of commercial tuning fork AFM sensors instead of standard cantilevers, as tuning fork sensors do not require optical readout. Specifically, we use Akiyama probes, which are both self-sensing and self-actuating, enabling fully electronic readout with only two wires.[40] A schematic of this probe is shown in Figure 1c. It operates by coupling the vibration of a piezoelectric tuning fork to a thin SiN cantilever mounted between its prongs. The resulting resonance modes can be used for topographic sensing. We employ the first cantilever bending mode, shown in Figure 1e, to perform both non-contact and tapping-mode AFM.[41] This mode keeps the tuning fork off resonance, which lowers the mechanical coupling and stabilizes the tapping mode.

Utilizing tapping-mode (also called dynamic-mode) AFM, we close the gap between the nanoSQUID and the surface. In this mode, the topographic feedback is determined by the repulsive part of the tip-sample interaction potential.[42] This ensures that the spatial resolution of the SOT is set only by the nanoSQUID size and oscillation amplitude.

A key feature of the nanoSQUID sensors in our probes is the use of proximity Josephson junctions with a normal-metal barrier (SNS junctions), formed by a ~20 nm long Cu nanobridge exposed through focused ion beam (FIB) milling. The major advantage of using fully metallic proximity weak links, compared to other types of Josephson barriers, is their non-hysteretic current-voltage characteristics at all temperatures. In contrast, traditional SOT and SOL probes use (constriction) Dayem bridges, which are known to be hysteretic, especially far below the superconducting transition temperature.[43–45] In addition, our Nb-Cu nanoSQUIDs have an exceptional transfer function of a few mV/$\Phi_0$ (see Supporting Information), which is at least an order of magnitude higher than the typical values reported for nanoSQUIDs.[46] Our junctions have a single-valued current phase relation at all temperatures, and maintain their characteristics in the mK regime.[47,48]

Utilizing multilayer heterostructures, we create a probe platform that enables flexible sensor customization. The versatility of the direct-write approach implemented here allows us to fabricate sensors ranging from 50 nm to several microns, depending on the desired trade-off between field sensitivity and resolution. The geometry of the device is easily modified, allowing us to tune SQUID asymmetry and thus its zero-field sensitivity. This is of particular interest to the study of spontaneous time-reversal symmetry breaking and field-sensitive quantum devices.

The supercurrent interference pattern of a typical AFM-SQUID probe is shown in Figure 1d. The large gradient in differential resistance, the deep modulation of the pattern and the substantial $I_cR_N$-product all contribute to the exceptional transfer function of our nanoSQUIDs. Due to the resulting large voltage, cryogenic amplifiers are no longer strictly necessary. All the measurements presented here were recorded in current-bias mode, using only a lock-in amplifier. This greatly simplifies our electronic setup and streamlines the operation of the microscope.

The experiments were performed inside a continuous-flow cryostat with the pulse tube running. We observe no interference from the pulse tube in our experiments. Even after seven weeks of continuous imaging in tapping-mode, neither the probe nor the sample showed any signs of degradation.

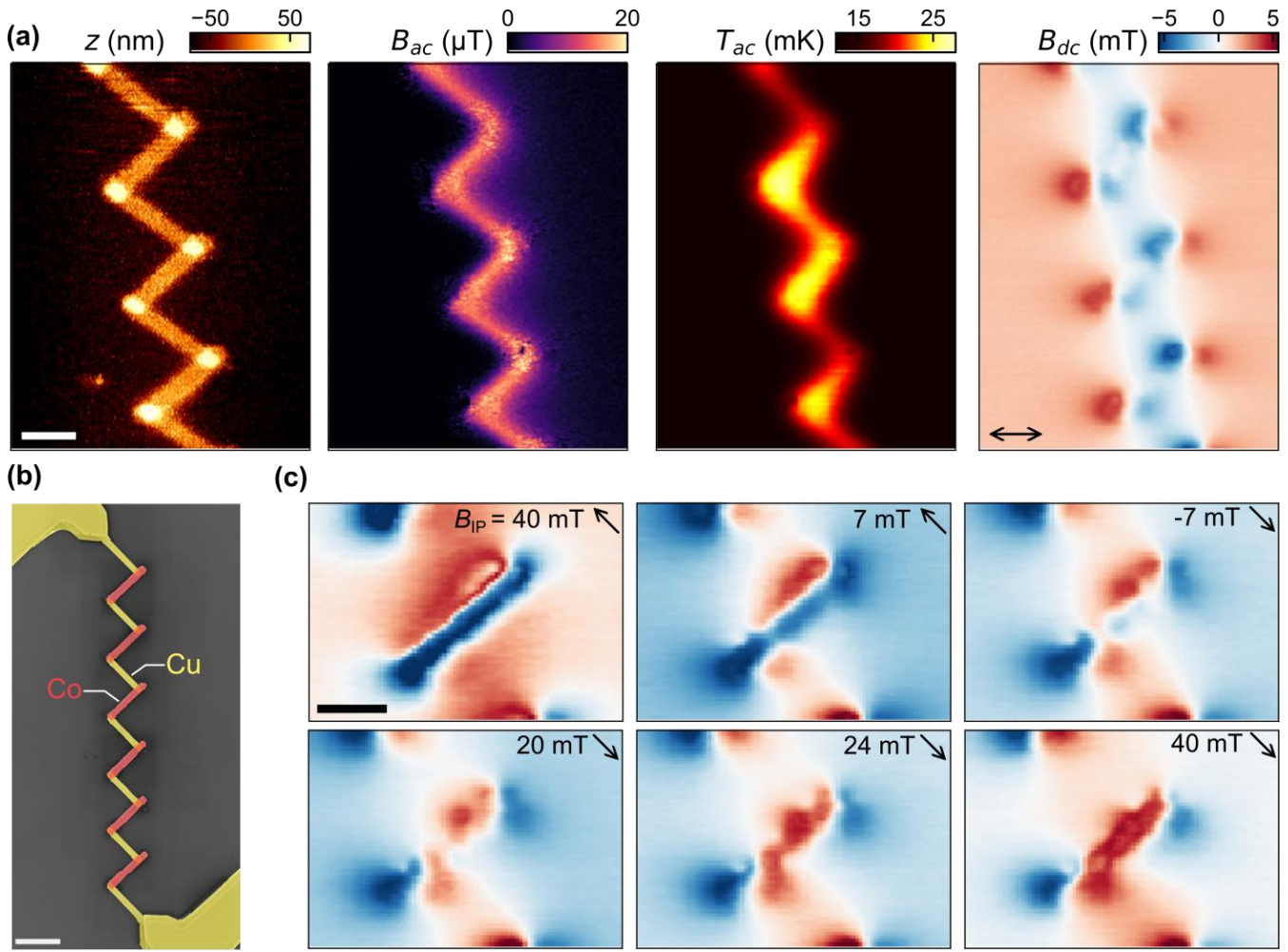

**Figure 2.** SQUID-on-tip images of a copper-cobalt heterostructure. **(a)** Simultaneous images of topography, current, dissipation and magnetism, while exciting the heterostructure with an a.c. current of 100 μArms. The scalebar denotes 2 μm. The arrow in the rightmost panel denotes the (sensitive) SQUID axis. **(b)** Scanning electron micrograph of the sample. The scalebar denotes 5 μm. **(c)** A single cobalt micromagnet in various perpendicular fields, highlighting domain-wall formation. The arrow in the panel denotes in-plane field size and direction. Using our vector-magnet, we simultaneously apply out-of-plane fields between –40 mT and +40 mT so that the probe remains at its most sensitive point during every scan. The scalebar denotes 2 μm. The total range of all images is 8 mT.

We demonstrate the working principle of the microscope with a magnetic heterostructure, consisting of alternating cobalt and copper nanowires, shown in Figure 2b. Due to the shape anisotropy, the cobalt is naturally magnetized along its long axis, creating dipole fields that can be imaged with our SQUID. Applying a current through the sample generates magnetic fields as well as dissipation from Joule heating.

Using an alternating-current bias, we demonstrate simultaneous imaging of all relevant parameters. The current generates magnetic fields at a frequency $f$, while the Joule heating oscillates at $2f$. The dipole fields are static and can therefore be measured via the SQUID's d.c. voltage. This frequency multiplexing method was first used by Halbertal *et al*.[30]

We simultaneously record the topography, static magnetic fields, current-induced a.c. fields, and the resulting heat, as shown in Figure 2a. Although the a.c. and d.c. signals differ by more than two orders of magnitude, the frequency-separation technique provides images which are nearly completely artifact-free. We obtain quantitative measurements using a simple one-to-one calibration (see Supporting Information). Comparing against the topographic image, we see that both transport signals are confined to the nanowire, underscoring the probe's utility as a local current sensor. The thermal image shows a contrast between cobalt and copper segments, due to their resistivity mismatch. The increased dissipation at the wire ends may indicate finite interface resistance.

Because our SQUIDs respond to both in-plane and out-of-plane fields, they can be used at optimal sensitivity over a large and continuous range of background fields. We illustrate this with an experiment in Figure 2c, where we image the magnetization reversal of the cobalt wires while sweeping an in-plane field. Initially, we magnetize the cobalt wires by applying a 200 mT field along their short axis, which is then incrementally inverted. We image the capture and gradual propagation of a domain wall, leading to magnetization reversal in the bar. Since the magnetization of cobalt nanostructures is insensitive to small out-of-plane fields, we use an out-of-plane field to keep the SQUID at its working point at all times. This additional degree of freedom is available because of the 62° angle between the SQUID and the surface. We proceed by benchmarking the unique capabilities of the microscope as a local probe for high-resolution transport phenomena. To characterize its current imaging capacity, we study a high-density niobium serpentine nanostructure, shown in Figure 3a.

In Figure 3b, we present a magnetic image of the serpentine driven by an a.c. current of 100 μA rms. Our in-plane sensitivity reveals the local current distribution, including clear sign inversions where the current reverses direction. This level of detail is inaccessible using conventional out-of-plane techniques.

We assess the probe's sensitivity by performing line scans at various applied currents. Figure 3c shows the raw signals at 10 μA and 1 μA without any integration or post-processing. Both signals exhibit good signal-to-noise ratio (SNR) and low field noise.

Subsequently, we demonstrate the probe's ability to resolve sub-μA currents by reducing the current to 100 nA. Our ability to simultaneously record topography is vital here, as it allows pixel-perfect alignment of multiple line scans, which can be integrated to obtain a high-SNR trace. This represents a significant improvement over previous out-of-plane imaging works, none of which could reach such low currents at sub-μm spatial resolution.[36,49–51]

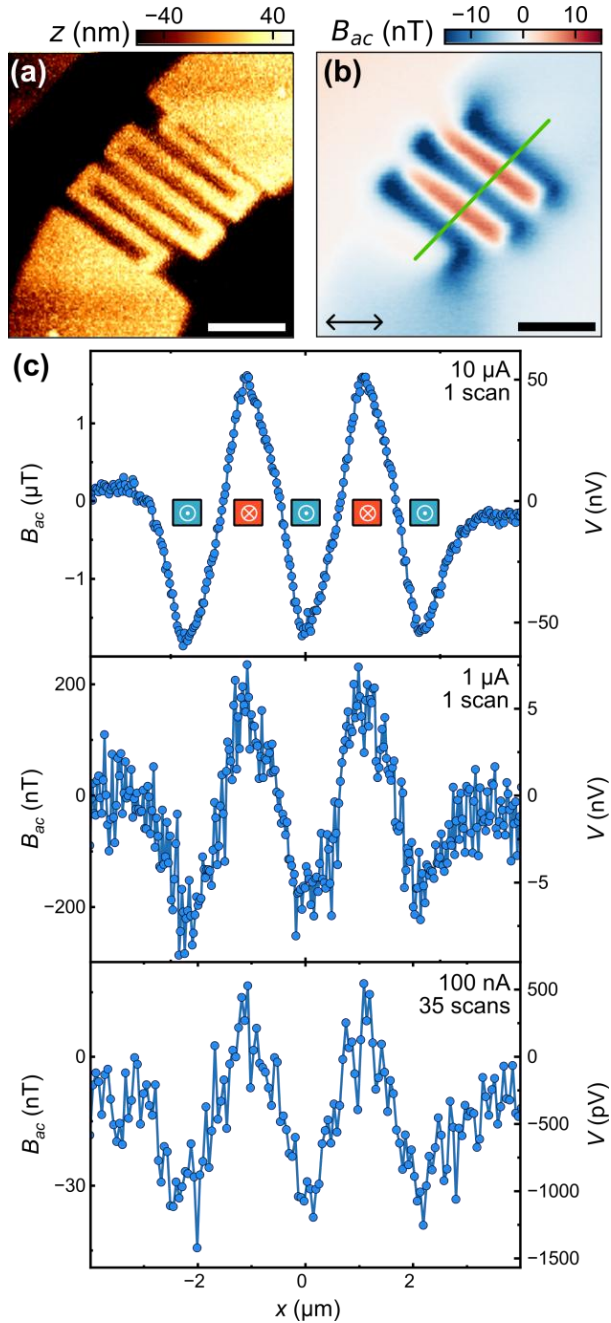

**Figure 3.** Nanoresolved imaging of small supercurrents **(a)** Topographic image of the Nb serpentine. **(b)** The magnetic field generated by a 100 µArms current. Arrow indicates SQUID axis. **(c)** Linescans taken along the green line in figure (b), each taken at a different excitation current, using lock-in time constants of respectively 100, 150 and 750 ms. Drawing in top panel shows the position of the wires and current direction. All scalebars: 2 µm.

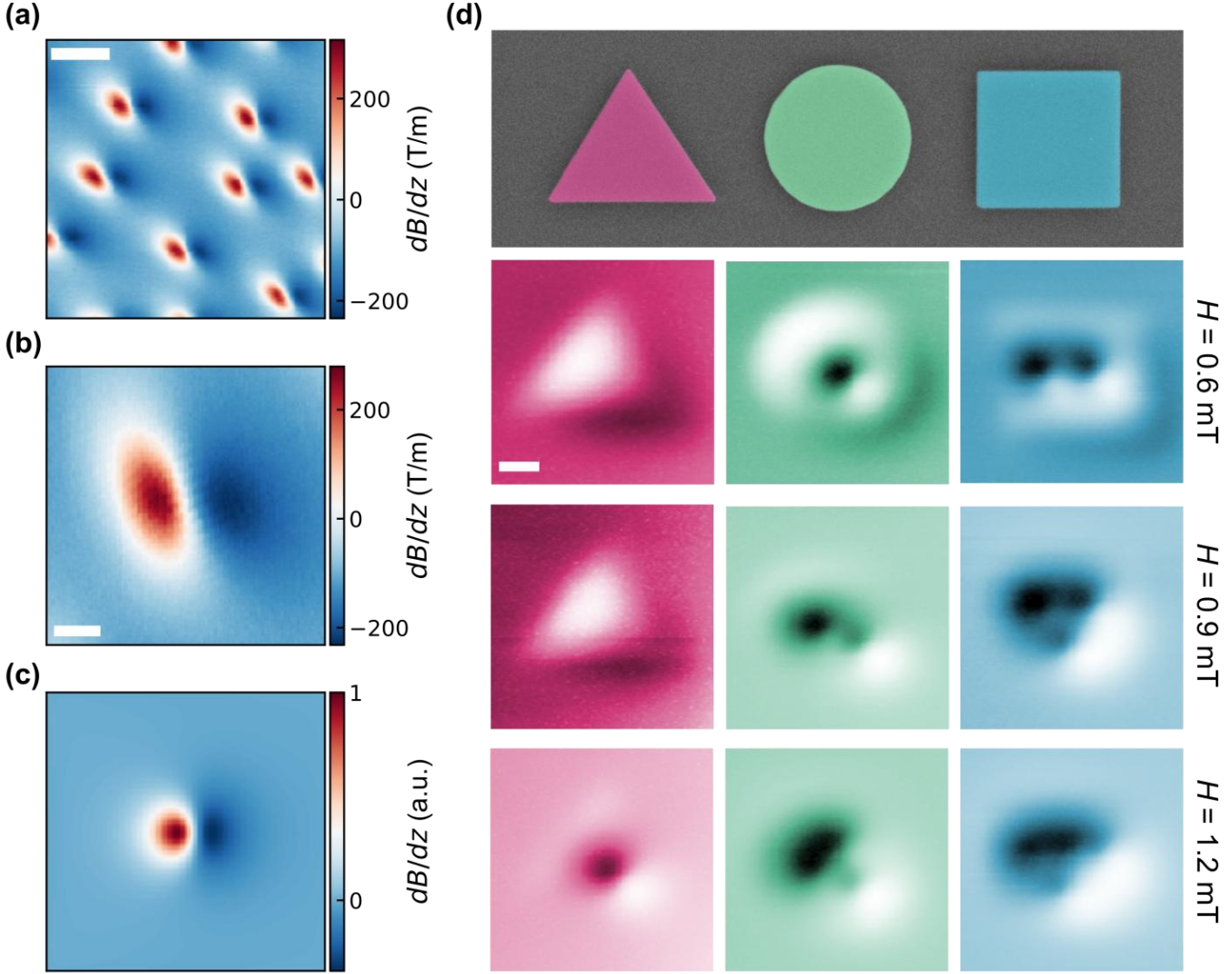

**Figure 4.** Vortex imaging with our SQUID-on-tip in its gradiometric imaging mode, where the sample is oscillated out-of-plane with 35 nm rms. Instead of imaging the vortex core, we observe the currents flowing around it. **(a-b)** Pearl vortices in a Nb thin film. These images are taken in non-contact AFM mode at a constant height of 250 nm. (b) shows a close-up of a single vortex in the same thin film with scalebar 1 μm. **(c)** Simulation of a single vortex for $\Lambda$ = 570 nm, which corresponds to the expected Pearl length of our film at 3.1 K. **(d)** Imaging vortices in microstructures using tapping-mode gradiometry. The top image shows a scanning electron micrograph of a triangle (side length of 5.3 μm), circle (diameter of 4.6 μm) and square (side length of 4.5 μm). These are cooled down in a finite field, denoted to the right of the images. Various fields yield different vorticities, depending on the edge-to-surface ratio. The scale of the images is, from top to bottom, for the triangles 500 T/m, 300 T/m and 800 T/m, for the circles 750 T/m, 1400 T/m, and 1800 T/m, and for the squares 1000 T/m, 1400 T/m and 2100 T/m. The scalebar in (a) is 4 μm, and 1 μm in (b-d).

We now describe how our technique can also be applied to static fields, such as those generated by magnetic materials and spontaneous supercurrents. The detection of small currents is typically done using a.c. techniques, to avoid 1/f noise, but these methods are not applicable to static fields. In our microscope, we address this problem by lightly oscillating the sample in the out-of-plane direction, thereby converting static signals to an a.c. flux. Similar methods were previously used in scanning NV,[52] SOT,[31] and SOL.[37]

Using this gradiometric imaging, we investigate the supercurrents flowing around Pearl vortices, which are the thin-film counterpart of Abrikosov vortices. In Pearl vortices, the supercurrent is expected to decay over the Pearl length $\Lambda = 2\lambda^2/d$, where $\lambda$ is the London penetration depth and $d$ is the film thickness.

In Figure 4a, we show images of randomly distributed vortices in a 0.1 mT field-cooled, 60-nm-thick Nb film ($T_c$ = 8.8 K). This arrangement of vortices is a product of the large inter-vortex spacing and the natural pinning sites present in the film. In this case, pinning forces dominate over vortex–vortex repulsion, preventing the formation of an Abrikosov lattice.

Notably, our microscope primarily images the circulating supercurrent around the vortex core, rather than the flux penetrating through it. This is illustrated in Figures 4b and 4c, which show experimental data and a simulation of a single vortex, respectively. The details of the simulation are provided in the Supporting Information.

Next, we study vortices in confined geometries: a triangle, circle and square microstructured from a 70-nm Nb film. We cool the structures below the transition temperature while applying small out-of-plane fields, ranging from 0.6 mT to 1.2 mT, and image the emerging vortices.

At 0.6 mT, a single vortex is present in the circle, and two vortices are present in the square. In the triangle, however, a vortex appears only when applying 1.2 mT. At this field, the circle and square host three and four vortices, respectively. We emphasize that the system is fully deterministic: experiments are repeatable over multiple thermal cycles, and fields below 0.2 mT result in a vortex-free state.

The differences in the nucleation fields can be understood through the potential energy landscapes of the three geometries. In thin-film nanostructures, vortex entry is governed by the Gibbs free energy, where the kinetic energy of the vortices competes against the vortex-Meissner repulsion. This interaction pushes vortices away from the edges and toward the center of a microstructure.[53] As the triangle has a larger perimeter-to-area ratio, it has the highest Gibbs barrier and requires a higher field for the first vortex to enter. Note that when the first vortex enters the triangle at 1.2 mT, the enclosed flux is already 7.2 $\Phi_0$. In contrast, in the circle the first vortex enters at 4.9 $\Phi_0$ and the square hosts two vortices at 6.0 $\Phi_0$.

In summary, we have introduced tapping-mode SOT, which utilizes a non-invasive electronic readout to simultaneously probe topography, magnetism, transport and dissipation. We benchmarked the microscope in magnetic hybrids and superconducting devices, where we demonstrate how in-plane field sensitivity and the large transfer function of the proximity nanoSQUID enable us to resolve currents down to 100 nA. The ultra-high sensitivity of our probes to magnetic, thermal and transport phenomena makes this an ideal technique for studying a wide range of quantum systems, both in exotic matter and sensitive quantum devices.

## AUTHOR INFORMATION

**Corresponding Author**

* Kaveh Lahabi, Huygens-Kamerlingh Onnes Laboratory, Leiden University, P.O. Box 9504, 2300 RA Leiden, The Netherlands; Email: lahabi@physics.leidenuniv.nl

## ACKNOWLEDGMENT

The authors would like thank Dalal Benali, Christiaan Pen, Peter van Veldhuizen, Fabian Steinmeyer and Maurits Geenen for their technical contributions, and Johannes Jobst for helpful discussions. This work was financed by the Dutch National Growth Fund (NGF), as part of the Quantum Delta NL Programme in the Project SME_R027_MQC, and the Dutch Research Council, as part of the "The Full Picture: Novel Microscopy with Smart Quantum Probes" (Project: VI.Veni.212.302), "Three seeds for the quantum/nano-revolution" (Project: NWA 1418.22.001), and Take-off Phase 2 (Project: 20964). It was also partly supported by European Union via the NextGenerationEU program.